%% file: main.tex
\documentclass[graybox]{styles/svmult}

\usepackage{mathptmx}       
\usepackage{helvet}         
\usepackage{courier}        
\usepackage{type1cm}        
\usepackage{makeidx}         
\usepackage{graphicx}        
\usepackage{multicol}        
\usepackage[bottom]{footmisc}
\usepackage{subcaption}

\makeindex             

\begin{document}

\title*{Kubernetes in the Cloud vs. Bare Metal: A Comparative Study of Network Costs}
\author{Rodrigo Mompo Redoli and Amjad Ullah}
\institute{Rodrigo Mompo Redoli \at Edinburgh Napier University, \email{rodrigo@rodrigomompo.com}
\and Amjad Ullah \at Edinburgh Napier University \email{a.ullah@napier.ac.uk}}
%
%
\maketitle

\abstract{Modern cloud-native applications increasingly utilise managed cloud services and containerisation technologies, such as Kubernetes, to achieve rapid time-to-market and scalable deployments. Organisations must consider various factors, including cost implications when deciding on a hosting platform for containerised applications as the usage grows. An emerging discipline called FinOps combines financial management and cloud operations to optimise costs in cloud-based applications. While prior research has explored system-level optimisation strategies for cost and resource efficiency in containerized systems, analysing network costs in Kubernetes clusters remains underexplored. This paper investigates the network usage and cost implications of containerised applications running on Kubernetes clusters. Using a methodology that combines measurement analysis, experimentation, and cost modelling, we aim to provide organisations with actionable insights into network cost optimisation. Our findings highlight key considerations for analysing network expenditures and evaluating the potential cost benefits of deploying applications on cloud providers. Overall, this paper contributes to the emerging FinOps discipline by addressing the financial and operational aspects of managing network costs in cloud-native environments.}

\input{introduction}
\input{relatedwork}
\input{methodology}
\input{experiments}
\input{discussion}
\input{conclusion}

%
%
\bibliographystyle{abbrv}
\bibliography{ref.bib}
\end{document}

%% file: introduction.tex
\section{Introduction}\label{secIntroduction}
The containerisation of applications and their deployment through orchestration tools like Kubernetes have significantly transformed the deployment and management of applications in cloud environments~\cite{ullah2021micado,ullah2023orchestration}. These tools facilitate efficient resource allocation and utilisation, resulting in cost savings and enhanced scalability. While many studies have examined system-level optimization for cost and resource efficiency in containerized applications \cite{mustyala2021dynamic, carrion2022kubernetes, ullah2017towards, ullah2018control}, few have focused on the specific cost implications of network usage in Kubernetes clusters within cloud environments. Resources in cloud environments can be quickly provisioned and de-provisioned on demand. This establishes a clear relationship between their usage and cost, making the cloud model attractive. However, to maintain application performance during traffic surges, network resources must be overprovisioned, which can lead to increased costs. Consequently, analysing and optimising network costs become challenging due to several factors such as burst traffic patterns, peak usage demands, data transfer rates, and inter-container communication~\cite{sfiligoi2021managing, ferdaus2017algorithm}.

Traditionally, bare-metal network links are priced by capacity, while managed cloud services charge based on actual usage. To prevent resource shortages, cloud providers use advanced provisioning strategies for servers and storage~\cite{shen2022resource, singh2016cloud, chen2018survey, ullah2018control}. However, implementing a pay-per-data-transfer model for network usage adds complexity, as providers must overprovision network links to ensure high availability during demand fluctuations. Providers have designed pricing models that factor in data transfer and network usage, including the overhead for dynamic scalability and application availability. This paper examines network costs for applications running on traditional bare-metal infrastructure versus managed cloud platforms with a usage-based model. The suitability of such a cost model for a particular application depends on its unique characteristics. However, understanding network usage patterns and scalability requirements is essential for organisations to make informed decisions when selecting hosting providers for containerised applications. 

This paper presents a methodology for analysing network costs, helping organisations assess the feasibility and potential savings of deploying containerised applications across various providers and pricing models. Specifically, it compares managed cloud providers---charging higher rates for outbound traffic---with bare-metal providers that price network links based on capacity. To address this, we performed a detailed cost analysis of a network-intensive application subject to higher outbound traffic costs on a managed cloud versus the equivalent on a bare metal. Our primary research question explores running a network-intensive application in the two enviornments regarding its cost-effectiveness. 

The rest of this paper is structured as follows: Section~\ref{sec:RelatedWork} reviews related work, Section~\ref{secMethodology} details the methodology, and Section~\ref{secExp} presents the experimental setup and results. Section~\ref{sec:discussion} discusses the findings and their implications, while Section~\ref{sec:conclusion} concludes the paper.

%% file: relatedwork.tex
\section{Related work} \label{sec:RelatedWork}
Major cloud providers offer tools to help organisations estimate and compare application hosting costs. For example, AWS Cost Explorer~\footnote{https://aws.amazon.com/aws-cost-management/aws-cost-explorer/}, Google Cloud Billing~\footnote{https://cloud.google.com/billing/docs?hl=es-419}, and Azure Cost Management~\footnote{https://azure.microsoft.com/es-es/products/cost-management} provide insights into cost breakdowns for their services. Although these tools provide valuable information on the overall cost analysis, they do not offer detailed network cost breakdowns. Especially when we have Kubernetes clusters hosting multiple applications, these tools only provide aggregated network costs and do not allow data segregation by application. Hence, making it difficult to identify subsystems that could benefit from a bare-metal approach.

Third-party tools also specialise in cost and network analysis for managed cloud environments, offering advanced features for comparing network costs across cloud providers and bare-metal hosting. They usually cover all aspects of a cloud provider, including support for Kubernetes cost analysis. Some examples of such tools include Cloudability~\footnote{https://www.apptio.com/products/cloudability/} and Archera~\footnote{https://archera.ai/}. These tools help organisations assess and optimise network costs, thus aiding in the decision-making process. Another popular tool, Datadog~\footnote{https://www.datadoghq.com/}---a market leader in monitoring---is potent for container resource allocation optimisation. However, none of these tools is open source. Furthermore, none of these consider the specific network requirements and costs associated with running applications in Kubernetes clusters except Datadog. In contrast, we propose using open-source Kubecost to analyse the network costs associated with running applications in Kubernetes clusters. Marino et al~\cite{marino2023dynamic} also use Kubecost to do a similar cost analysis, however, their research focuses on high-performance computing instead of network-intensive applications. 

There are also several research efforts, where resource management schemes are proposed to address the issues of network utilisation and cost in different environments. For example, some studies have focused on optimising resource allocation and load-balancing techniques in cloud environments to reduce network costs~\cite{verreydt2019leveraging, gao2020hierarchical, chhabra2021dynamic}. Others have developed cost estimation and prediction models to help organisations make informed decisions about hosting their applications~\cite{dong2023agent, cho2020cost, xu2018cost}. However, none of these approaches cover comparative aspects of bare metal vs managed cloud. Lastly, some research efforts, e.g. de Vries et al.~\cite{de2023cost}, gain insight into the costs of a particular application using performance metrics. However, application performance metrics need to be tightly integrated with the application, which may not be suitable for analysing an existing production cluster with multiple applications running that do not have application performance metrics integrated into their code. Tools and techniques that provide a holistic view of network usage and costs based on infrastructure metrics are needed in such cases. This paper aims to fill this gap by using infrastructure cost analytics tools to measure any application running in Kubernetes, which will allow organisations to apply FinOps methodologies later to select the best approaches to optimise costs.

%% file: methodology.tex
\section{Methodology} \label{secMethodology}
Figure~\ref{fig:methodology} depicts our methodology. The detailed description of each of the steps is as follows:
\begin{figure}[b]
    \centering
  \includegraphics[width=0.7\linewidth]{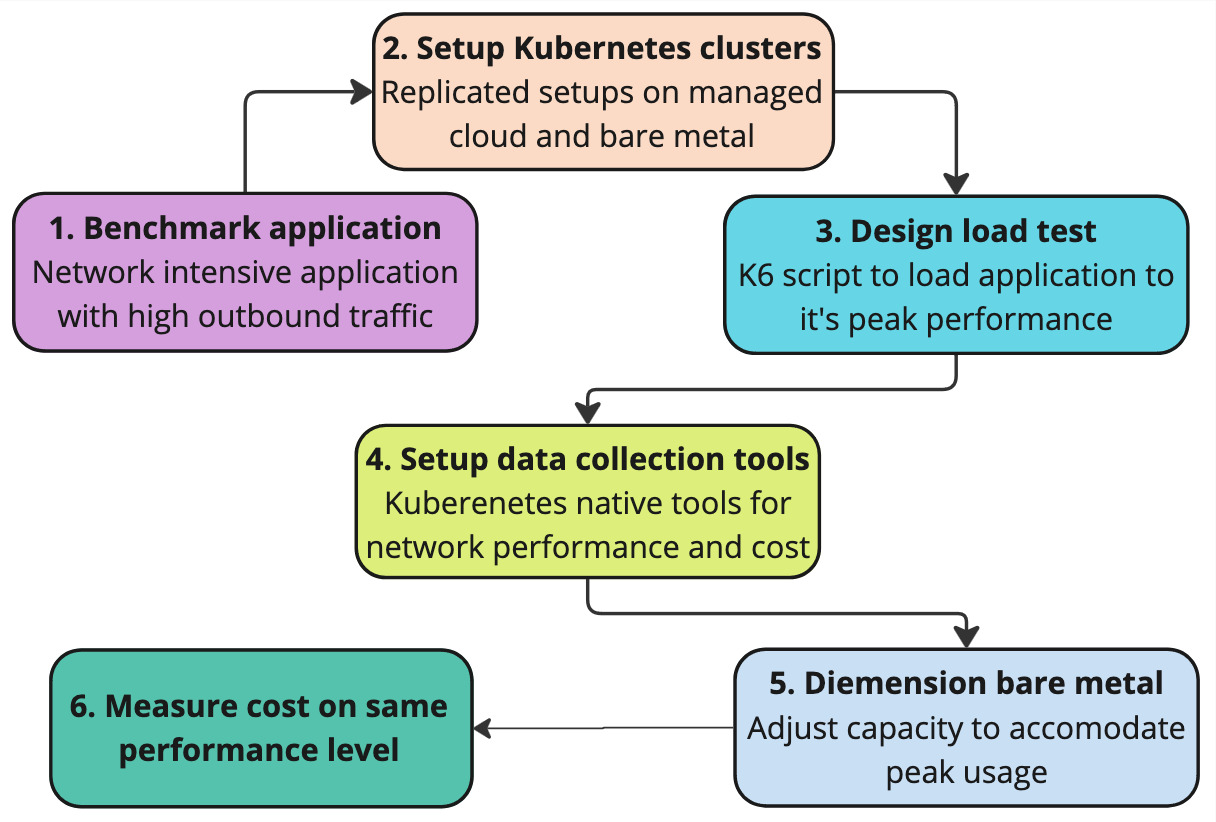}
  \caption{Overall methodology}
  \label{fig:methodology}
\end{figure}
\begin{enumerate}
    \item \textbf{Benchmarking application:}
Based on the recent annual Internet Report by CISCO~\cite{ciscoannualreport2023}, most of the worldwide traffic is related to video content with a continuously increasing trend in the future. Hence, a video processing or delivery-related application is a natural fit. Alternatively, applications with huge data uploads to cloud storage can also be a candidate. However, in such cases, most managed providers offer discounted inbound traffic prices, even free in some cases, e.g., AWS S3 storage. Therefore, we need to focus on applications with higher outbound traffic than inbound traffic where there is potential for reduced costs in bare metal. As a result, we selected a video delivery application instead of a video processing application, as video processing mainly involves inbound traffic or, in the best case, the traffic will be symmetric. We have chosen Plex Media Server~\footnote{ Plex: https://www.plex.tv/} for video delivery.

    \item \textbf{Setup Kubernetes clusters:} This step involves setting up identical Kubernetes clusters in both enviornments, i.e. on a bare metal and a managed cloud. To set up a Kubernetes cluster on a managed cloud, the managed Kubernetes solutions can be directly utilised. For example, we used EKS\footnote{ AWS EKS: https://aws.amazon.com/es/eks/}--- a managed Kubernetes cluster service provided by Amazon---for setting up the Kubernetes cluster on the AWS cloud. In the case of a bare metal, we utilised, K3d~\footnote{k3d:https://k3d.io/stable/}---a lightweight but production-ready deployment solution---for setting up Kubernetes clusters on bare metal keeping all the Kubernetes functionality. On top of these Kubernetes clusters, several tools are required to be deployed to simulate and measure real-world scenarios for network cost analysis. Additionally, we will also need to install monitoring tools to collect relevant data and metrics on network usage.  
    \item \textbf{Load test on the experimental setup:}
As part of the experimental setup, we used K6\footnote{ K6: https://k6.io/} for load testing to simulate high volumes of user traffic and measure the performance and scalability of our application. K6 is an open-source tool supported by a large community and provides flexible scripting capabilities, making it well-suited for our needs. It enables recording of a user journey in a browser and executes it from different simulated clients simultaneously. To build a k6 script, we use the k6 browser server recorder extension, which enables us to record user interactions on the application and generate a script automatically. We simulate the user journey to log in to the platform and stream a video from the content library. Once we have the K6 script, we can run the script with different configurations to simulate the data transfer at peak usage. Using this peak transfer data, we can extrapolate data for both scenarios to compare the cost over a month.
     \item \textbf{Setup data collection tools:}
Prometheus\footnote{Prometheus: https://prometheus.io/}, in combination with Grafana\footnote{ Grafana: https://grafana.com/} dashboards, are the most popular open-source tools that are widely used across the industry for data collection within Kuberentes-based ecosystems. These tools provide powerful monitoring capabilities and a flexible querying language, making them suitable for collecting and analysing network usage data and other performance and cost metrics. To visualise the relevant metrics, we can use the Grafana community dashboards to analyse the data for the most common metrics, such as CPU usage, memory utilisation, and network traffic. With these tools, we can measure the network traffic generated by the application. However, as on bare metal, we are provisioning the links in advance; we will have a fixed fee for the network costs, independently of network usage. In the case of a managed cloud, there are additional considerations regarding network costs such as region/zone factors, communication between services, the role of the load balancer, etc. To get aggregated cost analytics, one tool that can be used for collecting cost-related data in a managed cloud environment is Kubecost. It provides visibility into cost allocation and analysis for Kubernetes clusters, including network costs.

\item \textbf{Dimensioning bare Metal networking:}
Managed cloud providers (e.g. AWS, Azure, etc) offer scalable infrastructure where resources, including network bandwidth, are charged per usage and scale automatically within set limits. These limits typically affect only large production systems, which may have customized pricing agreements. Therefore, most users do not need to pre-allocate bandwidth on the managed cloud. In contrast, running applications on bare-metal infrastructure requires careful resource planning based on measured data, as network bandwidth must be pre-allocated. Overloading pre-allocated links on bare metal can degrade performance or cause failures, necessitating strategic overprovisioning to balance cost efficiency and reliability. For our research, we assume that the physical network links are not optimised and there is a certain amount of wasted resources. For our experimentation, we use the CISCO guideline of 50\% network utilisation on the link at peak usage to ensure optimal performance and avoid network congestion despite increasing the costs~\cite{ciscobestpractices}. In real-case scenarios, the overprovisioning can be reduced by using traffic modelling~\cite{van2006network}. Considering this, if we purchase a bare metal server from a provider like OVHCloud, they charge \$147 per extra 1GBit/s. The bare metal server Advance 2 includes an 8-core and 16 threads processor with 32GB of RAM that provides for 1Gbps per instance for \$176.66. Please note that we do not consider any other cost in terms of CPU, memory, or storage, as the research is focused on the network only. Furthermore, the associated operational costs are also not considered as this varies across different providers.
\item \textbf{Measure cost on same performance level:}
The underlying network infrastructure on a managed cloud provider hugely differs from a bare metal cluster. A bare metal cluster has access directly to the physical network interfaces. In contrast, on a managed cloud provider, the network infrastructure is abstracted and provided as a service by the cloud provider. This abstraction and service-based model of network infrastructure on a managed cloud provider can bring advantages in terms of scalability, flexibility, and ease of management. However, it can also impact performance and change application behaviour. In the particular case of our application, the amount of data downloaded is not deterministic as the browser will adjust the amount of video downloaded to the local cache depending on the resource availability. For this reason, we must adapt the load test to ensure the amount of data transfer per second is equivalent in both environments to allow an accurate comparison of costs. We can use the data available on Grafana dashboards---visualises various metrics related to application performance and resource utilisation---to adjust the load test parameters. We do not consider hitting the maximum network capacity as the network automatically scales on a managed cloud. After adjusting the load test to generate the same amount of traffic in both enviornments, we can run the load test and record the cost measurements collected by Kubecost. 
\end{enumerate}

%% file: experiments.tex
\section{Experimentation and Results} \label{secExp}
\begin{figure}[t]
\centering
  \includegraphics[width=0.7\linewidth]{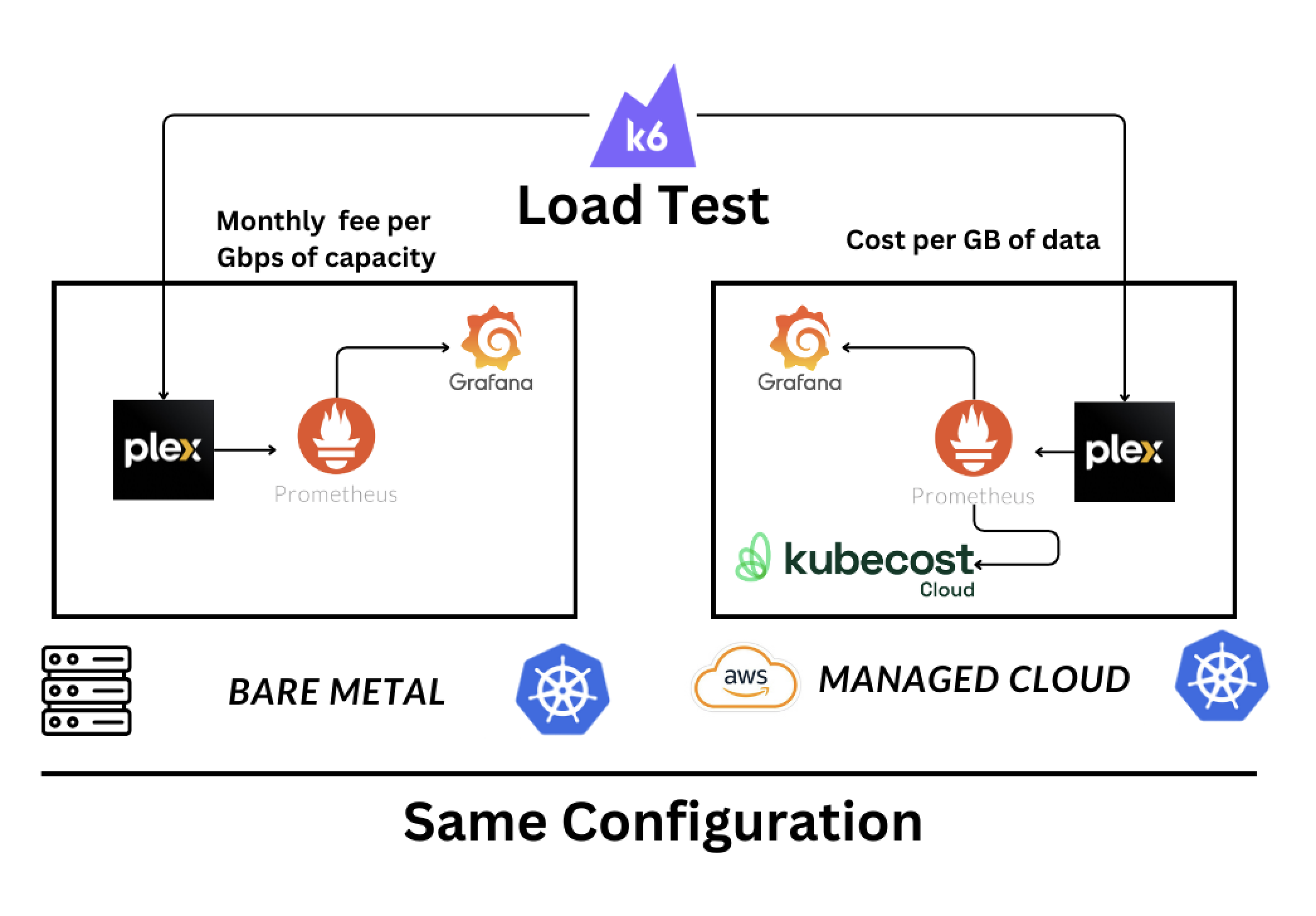}
  \caption{Experimental Setup with K6 on Bare Metal and Managed Cloud}
  \label{fig:expSetup}
\end{figure}

Figure~\ref{fig:expSetup} depicts the overall experimental setup to carry out the experimentation. The following subsections report the obtained results. 
\subsection{Identical performance in both enviornments}\label{sec:identicalPerformance}
\begin{figure*}[b]
    \centering
    \begin{subfigure}[t]{0.5\textwidth}
        \centering
        \includegraphics[width=\linewidth]{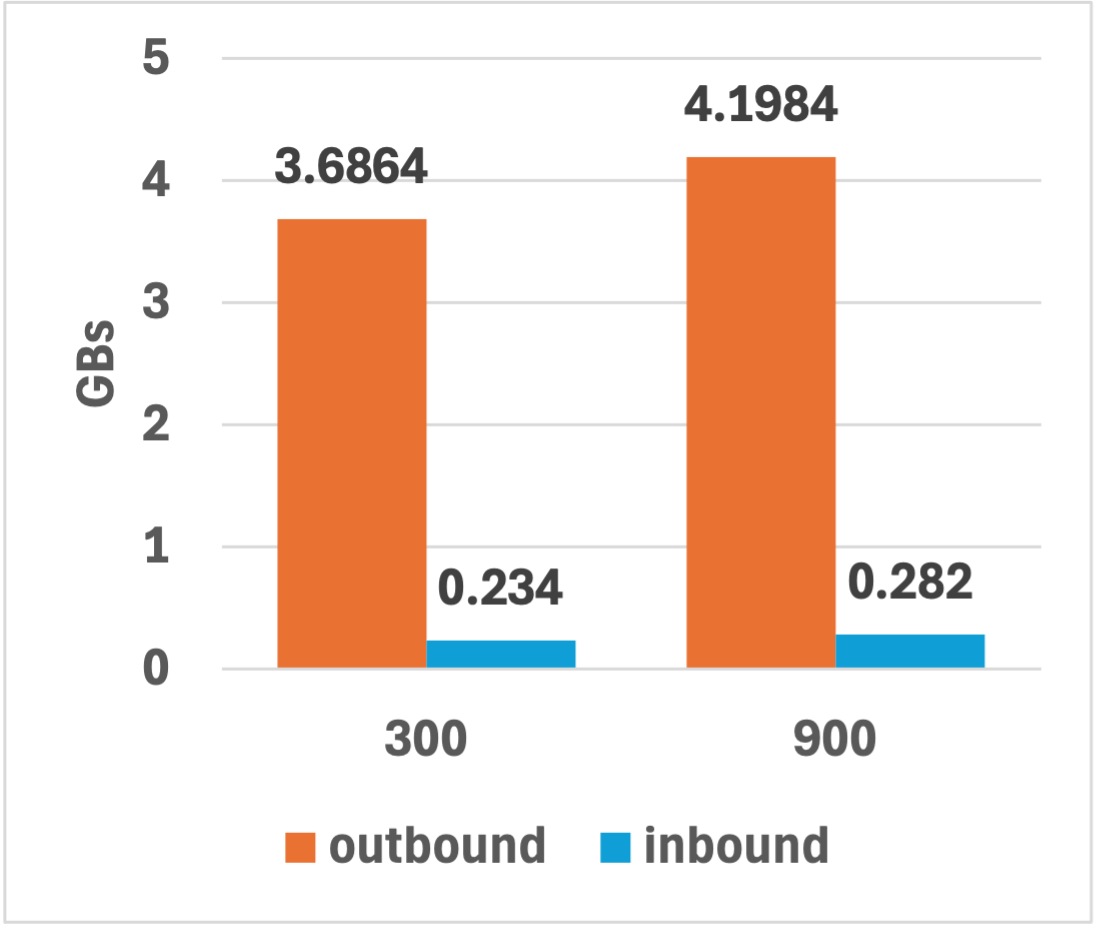}
        \caption{On bare metal with 300 and 900 users}
        \label{fig:loadtest_baremetal}
    \end{subfigure}%
    ~ 
    \begin{subfigure}[t]{0.5\textwidth}
        \centering
        \includegraphics[width=\linewidth]{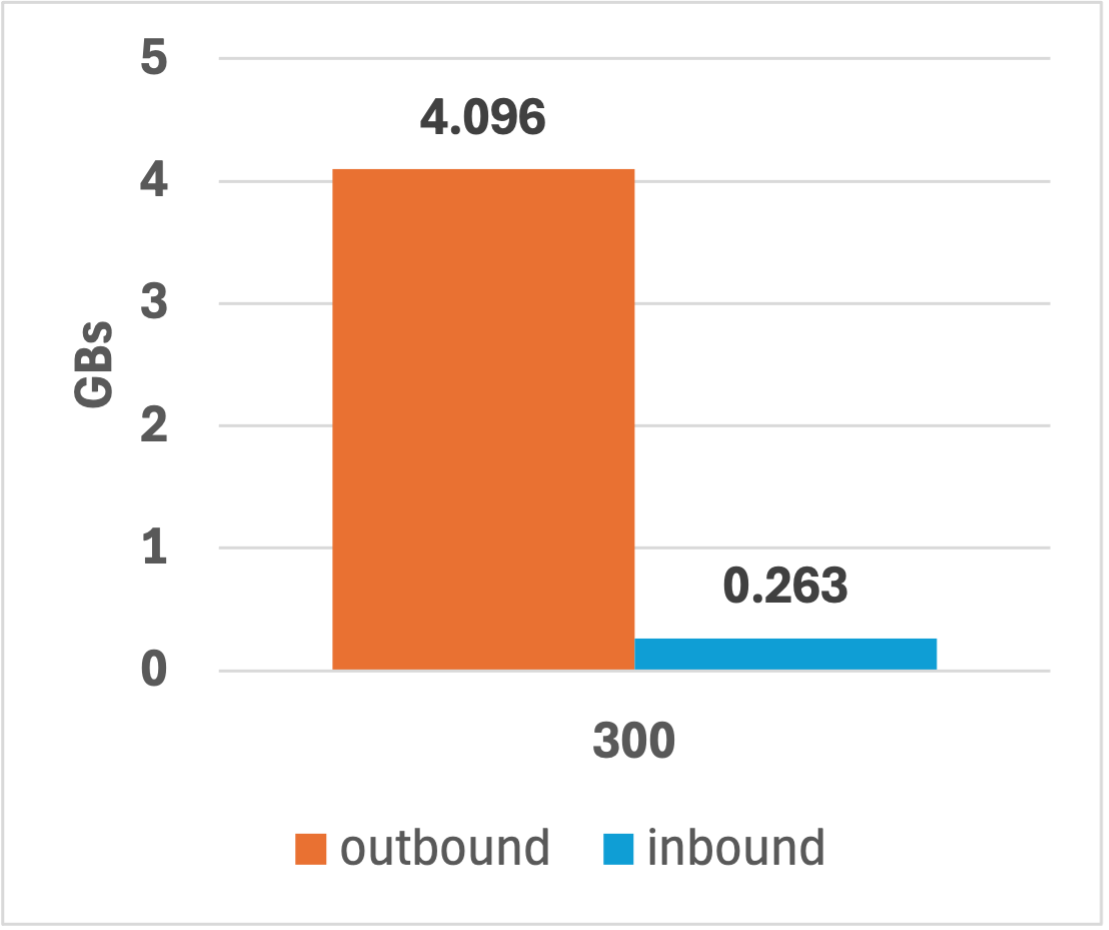}
        \caption{On managed cloud with 300 users}
        \label{fig:loadtest_managed}
    \end{subfigure}
    \caption{K6 load test results for 5 minutes}
\end{figure*}

\begin{figure}
\centering
  \includegraphics[width=0.9\linewidth]{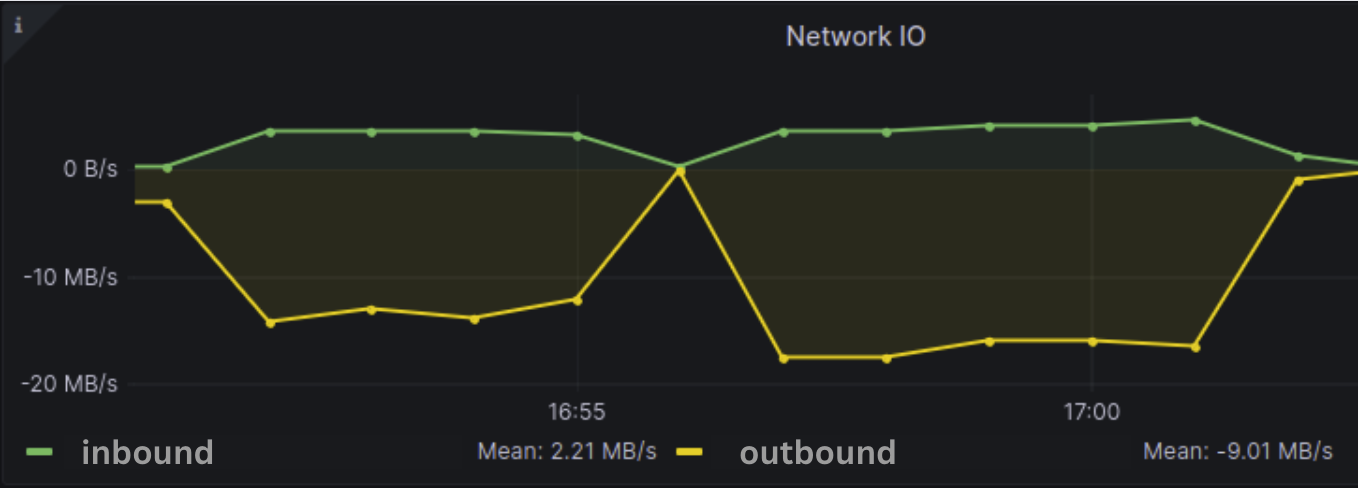}
  \caption{Bare Metal Pod Data Transfer with 300 Concurrent Users from 16:49 to 16:55 and 900 Concurrent Users from 16:56 to 17:02} 
  \label{fig:grafana_baremetal}
\end{figure}

For the evaluation, it is important to ensure that the amount of data transfer per second remains consistent across both environments, i.e. on bare metal and managed cloud. This consistency ensures that the comparison of costs is accurate and reliable. The bare metal load test achieved an average data transfer rate of 13 MBps with a total data transfer of nearly 4 GB over five minutes with 300 concurrent users (Figure~\ref{fig:loadtest_baremetal}). The transferred data remained well below the link capacity of 1 Gbps and below the target performance of the system. We further increased the number of concurrent users from 300 to 900 users. However, as is evident from Figure~\ref{fig:loadtest_baremetal}, adding more users did not result in significantly higher data transfer, indicating a system bottleneck, however, not related to network congestion. As a result, in this comparative analysis, we will conduct the cost comparison for a single Plex Media Server based on 300 concurrent users---the peak performance we obtain with our experimental setup. This measurement was obtained from a K6 client. We also checked this from inside the cluster using Grafana (Figure~\ref{fig:grafana_baremetal}) and found that the same amount of data was transferred from the Kubernetes cluster. Next, we performed the same load test on a managed cloud setup to confirm that the data transfer is equivalent to the bare metal setup. We observed a similar average data transfer rate, as shown in Figure~\ref{fig:loadtest_managed}. This verifies that both deployments perform at the same performance level.

\subsection{Kubecost configuration}
Kubecost allocates cost as a proportion of CPU usage. However, network costs are not proportional to CPU and RAM usage, therefore, the Kubecost network cost analyser needs to be configured to associate the correct amount of network usage to each service. Kubecost enables us to expand the total amount of the consumed traffic and get data about the traffic destinations from each pod. These destinations can be of the following types: in-zone, in-region, cross-region, and internet traffic. Each type of destination traffic incurs different pricing, and we need to factor that in for more accurate cost calculations. 

In-zone traffic refers to all the network traffic that occurs inside the same availability zone, typically located in the same building or a building very close to it by meters. In AWS, this traffic is usually free or comes at a low cost as it stays within the confines of the same physical infrastructure. Cross-zone traffic within the same region incurs a higher cost as it needs to traverse 100 kilometres through dedicated links maintained by the managed provider. In the case of AWS, this cost is typically 0.02\$ per GB of data, which is much cheaper than internet-egress traffic, which is charged at 0.09\$ per GB. The ingress traffic from the internet on the other hand is free. In the case of AWS, internet traffic is the only one without a symmetric price between ingress and egress. Cross-region traffic refers to network activity between different AWS regions. It involves data transfer across larger geographic distances, incurring higher costs due to increased distance and infrastructure involved, although still cheaper than internet traffic if configured correctly. 

 The different types of traffic---along with the different associated costs---make it difficult to calculate the real network costs. Furthermore, if the load-balancer ingress pod (e.g. NGINX in our case) is in a different zone, we will need to sum up the costs of cross-zone traffic, load-balancer costs, and internet traffic to get the total cost. Therefore, to obtain accurate measurements, we need to adjust our monitoring solution to adapt accordingly. Kubecost allows us to define to which zone the different subnets belong to ensure that traffic is tagged appropriately. Also, we can create the NGINX ingress with a static IP to tag all the traffic that reaches the NGINX ingress as internet traffic. In summary, we are using a mix of in-zone, in-region, and internet traffic to ensure that all traffic is billed at the correct rate. Since our application is deployed in a single region, hence no cross-region traffic. The complete configurations can be seen from our GitHub repository~\footnote{https://github.com/rodriwp/tfm/blob/main/argocd/apps-manifests/monitoring-kubecost/values.yaml}.  
\subsubsection{Network costs on managed cloud}
Using the above-mentioned configurations, we can obtain accurate network costs associated with running our application on the managed cloud. Table~\ref{tab:costs} presents the obtained results from a 30-minute experiment. It is important to note that all components of our application are currently allocated in the same region, which is considered ideal from a cost perspective; however, if there is a need for a cross-region service location for high availability, the costs will be slightly higher. Based on the cost of using the application over 30 minutes, we can extrapolate this cost to a monthly basis, assuming constant usage throughout, which corresponds to a constant network pattern. The total monthly cost would be around \$4800, which is significantly higher than the monthly cost of \$176.66 per month on a provider like OVHCloud for 1Gbps, independently of the traffic usage pattern. Compared to our bare metal scenario, we are increasing the network cost by 850\% on the highest utilisation scenario. However, it is important to note that the cost of running the application on a managed cloud may vary depending on the specific cloud provider and their pricing structure.
\begin{table}[b]
\begin{center}
    \setlength{\tabcolsep}{1em}
    \renewcommand{\arraystretch}{2}%
    \caption{Plex Media Server network costs for 30 minutes}
    \begin{tabular}{c|c|c|c|c} 
       & CPU & Memory & Persistent volume (PV) & Network\\
      \hline
      \textbf{Cost} & \$1.65 & \$1.94 & \$0.01 & \$3.34 \\
      \hline
    \end{tabular}
    \label{tab:costs}
\end{center}
\end{table}

%% file: discussion.tex
\section{Discussion} \label{sec:discussion}
Our primary research question was whether running a network-intensive application would be more cost-effective when deployed on a managed cloud or running it on a bare metal host provider. From the data we gathered based on the sample application, it is clear that the potential cost savings of running such an application on bare metal infrastructure are substantial, making it the cheaper option. However, when deciding between managed cloud or bare metal infrastructure, it is essential to consider other factors beyond just cost. For example, network-intensive applications, such as AI tagging or video streaming services without a regular usage pattern, may not necessarily benefit from a bare metal cost model with fixed network prices. In contrast to our selected benchmark application (Plex), we now comment on other network-intensive applications that could take advantage of the managed cloud cost model and potentially run cheaper on a managed cloud provider than on bare metal.

First, as ingress traffic is free on a managed cloud (e.g. on AWS), running the application on the managed cloud may be more cost-effective if the application is heavily dependent on ingress traffic. For example, an AI tagging video service in which the output of processing video is just metadata without the video. In this case, we would not benefit from the lower prices of bare metal as our primary usage is to ingest data. However, as the egress cost is significantly more expensive in AWS than in bare metal, if a small portion of the ingested videos is also delivered to another service for further processing, despite having more ingress traffic, the application could be cheaper running on bare metal. Therefore, we cannot make a generic decision without having concrete data on network usage like the one obtained with Kubecost.

Another interesting aspect, in addition to the ingress vs egress balance, is the network pattern of the application. Let's suppose we have an application with a highly variable traffic pattern in contrast to the stable amount of traffic as in the case of the benchmark application from this paper. An example would be a private streaming service without a CDN in which we only go live a couple of hours a day, and the core system components are powered off the rest of the time while there is no live streaming, taking advantage of the flexibility of the cloud. The costs will be significantly lower since egress data transfer is hugely minimized, as the system is utilized for only 8\% of the day. Using the above calculations, this will incur a monthly cost of \$384 on a managed cloud, which is still more expensive than running the application on bare metal, however, the cost difference between both hosting options is heavily reduced. 

Lastly, it is important to note that cost considerations alone should not be the sole determining factor in choosing between running an application on bare metal or a managed cloud host provider, as commercial agreements and opportunity costs may justify opting for a more expensive solution. Hence, we can not universally answer the question of whether it is cheaper to run the application on bare metal or a managed cloud from a business perspective. This research, however, provides a methodology to measure network costs in a consistent manner, which for network-intensive applications is one of the key cost indications due to its high impact on the overall costs. Having accurate data is crucial when making cost-related decisions in a data-driven manner, which is the main focus of FinOps professionals who optimize costs for technological businesses.

%% file: conclusion.tex
\section{Conclusion} \label{sec:conclusion}
Our analysis based on the benchmarking application shows significant network cost savings, making bare-metal hosting the more economical option. However, this may not apply to all network-intensive applications. In this research, we have identified several key aspects to be considered when answering that question for any other network-intensive application. Firstly, the nature of the application’s traffic pattern plays a crucial role in determining the cost-effectiveness of each infrastructure. Applications with highly variable traffic patterns may yield different cost outcomes compared to those with stable traffic. Moreover, we also highlighted that while egress traffic has a high cost on managed cloud providers like AWS, the ingress costs are free. This makes it essential to have a deep understanding of the specific traffic requirements and patterns of the application before making a decision. 

Overall, our study has provided a systematic methodology for measuring network costs at the same performance level, which is crucial for making informed decisions, especially for network-intensive applications. It is important to note that while cost savings may be a significant factor, the overall operational costs and other non-monetary considerations should be considered when deciding between managed cloud and bare metal infrastructure. Furthermore, leveraging a tool like Kubecost will allow the collection of accurate network data and cost data, which can be correlated with the application’s planned growth to understand at which point it will be cheaper to run on bare metal due to the high network costs, allowing teams to plan. For the next steps, our focus is mainly two-fold. First, broaden the cost comparison to include storage, development, and operational costs to have a comparison from a holistic viewpoint. Second, exploring the impact of different auto-scaling strategies on network usage and costs could provide further valuable insights. 